\documentclass[letterpaper, 10 pt, conference]{ieeeconf}  
\IEEEoverridecommandlockouts 

\usepackage{cite}
\let\labelindent\relax

\usepackage{xcolor}
\usepackage{epstopdf}
\usepackage{mathtools}
\usepackage{enumitem}
\usepackage{cancel}
\usepackage{fancyhdr}
\usepackage{subcaption}
\usepackage{amsfonts}
\usepackage{amssymb}
\usepackage[T1]{fontenc}
\usepackage{color}	
\usepackage{siunitx}
\usepackage{multirow}
\usepackage{placeins}
\usepackage{booktabs}
\usepackage{rotating}
\usepackage{array}
\usepackage[ruled]{algorithm2e}
\usepackage{algpseudocode}
\usepackage{physics}
\usepackage{dsfont}
\usepackage{bbm}
\usepackage{bbold}
\usepackage{caption}
\usepackage{subcaption}
\usepackage[hyphens]{url}

\usepackage{epsfig}
\usepackage{epstopdf}

\usepackage{xcolor}

\usepackage[prependcaption,colorinlistoftodos]{todonotes}

\newcommand{\mc}[1]{\mathcal{#1}}
\newcommand{\R}{\mathbb{R}}
\newcommand{\N}{\mathbb{N}}
\newcommand{\bs}[1]{\boldsymbol{#1}}

\DeclareMathOperator*{\argmin}{argmin}   
  
\usepackage{mathtools}
\usepackage[acronym]{glossaries}
\newacronym{GNE}{GNE}{generalized Nash equilibrium}
\newacronym{GNEP}{GNEP}{generalized Nash equilibrium problem}
\newacronym{vGNE}{v-GNE}{\textit{variational} generalized Nash equilibria}
\newacronym{KKT}{KKT}{Karush-Kuhn-Tucker}
\newacronym{LP}{LP}{linear program}

\newtheorem{theorem}{Theorem}
\newtheorem{lemma}{Lemma}
\newtheorem{definition}{Definition}
\newtheorem{assumption}{Assumption}
\newtheorem{stassumption}[assumption]{Standing Assumption}
\newtheorem{proposition}{Proposition}

\newtheorem{remark}{Remark}

\title{\LARGE \bf
Generalized uncertain Nash games: Reformulation and robust equilibrium seeking - Extended version}

\author{Marta Fochesato$^{\star}$, Filippo Fabiani$^{\dagger}$, John Lygeros$^{\star}$
\thanks{$^{\star}$ Automatic Control Laboratory, Department of Electrical Engineering and Information Technology,
        ETH Z\"urich, Physikstrasse 3 8092 Z\"urich, Switzerland
        ({\tt\footnotesize \{mfochesato, jlygeros\}@ethz.ch}).
}
\thanks{$^{\dagger}$ IMT School for Advanced Studies Lucca, Piazza San Francesco 19, 55100 Lucca, Italy ({\tt \footnotesize filippo.fabiani@imtlucca.it}).
}
}

\begin{document}
\maketitle
\thispagestyle{empty}
\pagestyle{empty}

\begin{abstract} 
We consider generalized Nash equilibrium problems (GNEPs) with linear coupling constraints affected by both local (i.e., agent-wise) and global (i.e., shared resources) disturbances taking values in polyhedral uncertainty sets. 
By  making use of traditional tools borrowed from robust optimization, for this class of problems we derive a tractable, finite-dimensional reformulation leading to a deterministic ``extended game", and we show that this latter still amounts to a GNEP featuring generalized Nash equilibria ``in the worst-case". 
We then design a fully-distributed, accelerated algorithm based on monotone operator theory, which enjoys convergence towards a Nash equilibrium of the original, uncertain game under weak structural assumptions. Finally, we illustrate the effectiveness of the proposed distributed scheme through numerical simulations.
\end{abstract}

\IEEEpeerreviewmaketitle

\section{Introduction}\label{sec:intro}
Game theory has recently attracted considerable research attention as a decision-making framework able to model problems involving heterogeneous agents that potentially share and compete for common resources. In this context, equilibrium notions characterizing generalized Nash equilibrium problems (GNEPs) \cite{GNEP:Facchinei} find broad applicability in social science and engineering, encompassing problems in power grid \cite{power}, traffic management \cite{traffic}, sensing and networks \cite{sensors}. 

While the largest portion of research efforts concentrate on GNEPs with deterministic data (i.e., agents' cost functions and local/coupling constraints), in many real-world applications the multi-agent system at hand may be strongly affected by uncertainty, thus possibly making standard solution concepts and algorithms ineffective. This motives us to look for tailored GNEP formulations able to handle uncertainties. Available approaches in robust game theory typically deal with uncertainty characterized by specific models of either the probability distribution characterizing the disturbance \cite{chance_constraint}, the geometry of the underlying support set \cite{robust} or, more recently, exploit the availability of historical disturbance realizations to propose data-driven distribution-free approaches \cite{scenario}. 
If, on the one hand, some recent works employed this framework to study GNEPs with uncertain cost functions (see \cite{Franci,Fele} just to mention a few), on the other hand the case involving uncertain constraints has been far less considered. Most notably, \cite{Peng} considers structured local uncertainties in the coupling constraints and proposes a distributed, continuous-time algorithm for seeking a $\varepsilon$-GNE. However, it does not consider uncertainties affecting the right-hand side, namely the vector of shared resources, which are typically encountered in real-world settings (especially in power network games \cite{mio}). Instead, these latter have been thoroughly investigated in \cite{Paccagnan,Filippo,fabiani2022probabilistic,Pantazis} under the lens of the scenario approach to quantify the robustness of the resulting equilibria against unseen realizations of the random variable.

Along this research direction, we consider GNEPs with linear coupling constraints affected by both local (i.e., agent-wise) and global (i.e., in the shared resources) disturbances taking values in polyhedral uncertainty sets, for which we propose an ``extended game", deterministic reformulation leveraging traditional tools borrowed from robust optimization, and a fully-distributed, accelerated equilibrium seeking algorithm. We hence summarize our contributions as follows:
\begin{enumerate}
    \item We provide a finite-dimensional, worst-case reformulation of otherwise intractable GNEPs with uncertain linear coupling constraints, and we show that the resulting ``extended game'' inherits that same structural properties of the original problem (\S \ref{sec:reformulation}).
    \item For the resulting deterministic GNEP, we prove that a relaxed-intertial scheme, leading to a fully-distributed algorithm over a graph, enjoys convergence guarantees to the set of equilibria of the original game (\S \ref{sec:algorithm}).
    \item Finally, we validate the proposed theoretical results numerically on an illustrative example (\S \ref{sec:numerics}).
\end{enumerate}
As discussed also later in the paper, we stress that the proposed worst case-based methodology is quite general and applies to a broad class of generalized games. 
Moreover, the proposed relaxed-intertial scheme converges under mere monotonicity of the game mapping, thus circumventing the assumption on the uniqueness of the equilibrium, a condition frequently encountered in algorithmic game theory \cite{notation,Pavel}.

The proofs of the technical results are all in Appendix.

\smallskip

\subsection{Preliminaries}
\subsubsection{Notation} $\mathbb{R}$, $\mathbb{R}_{\geq 0}$ and $\bar{\mathbb{R}} \coloneqq \mathbb{R} \cup \{\infty\}$ denote the set of real, nonnegative and extended real numbers respectively. 
Given $N$ vectors $x_1, \ldots, x_N \in \mathbb{R}^n$ and $\mathcal{I}\coloneqq \{1,\ldots, N\}$, we denote \mbox{$\bs{x} \coloneqq (x_1^\top,\ldots, x_N^\top)^\top = \textrm{col}((x_i)_{i \in \mathcal{I}})$}. The \mbox{$j$-th} element of a vector $v$ is denoted by $v_j$. Given a matrix $A\in\mathbb{R}^{m\times n}$, its transpose is denoted by $A^\top$, while $A \otimes B$ indicates the Kronecker product
between matrices $A$ and $B$. $ A \succ 0$ ($\succeq 0$) stands for a positive definite (semidefinite) matrix. For $A \succ 0$, we denote $\|\cdot\|_A$ the $A-$induced norm such that $\|x\|_A \coloneqq \sqrt{x^\top A x} = \sqrt{\langle Ax,x \rangle}$, where $\langle \cdot, \cdot \rangle : \mathbb{R}^n \times \mathbb{R}^n \rightarrow \mathbb{R}$ stands for the standard inner product.  $I_{n}\in\mathbb{R}^{n\times n}$, $\mathbf{1}_n \in\mathbb{R}^n$ ($\mathbf{0}_n$) $ \in\mathbb{R}^{n\times n}$ denote the identity matrix and the vector of all $1$ ($0$), respectively. 

\subsubsection{Operator theory} \label{prelim:OT}
Let $T : \mathbb{R}^n  \rightrightarrows \mathbb{R}^n$ be a set-valued operator.  The domain of $T$ is defined by $\text{dom}(T) = \{ x \in \mathbb{R}^n \:|\: T(x) \neq \emptyset \}$.
The set of zeros of $T$ is denoted as $\text{zer}(T) = \{ x \in \mathbb{R}^n \:|\: 0 \in T(x)  \}$. The set of fixed point of $T$ is denoted as $\text{fix}(T) = \{ x \in \mathbb{R}^n \:|\: x \in T(x)  \}$. An operator $T$ is monotone if $\langle T(x) - T(y), x - y \rangle \geq 0$ and it is ${\beta}$- Lipschitz continuous if $\|T(x) - T(y) \| \leq {\beta} \|x - y\|$. Given $T:\mathbb{R}^n \rightrightarrows \mathbb{R}^n$, the variational inequality problem $\text{VI}(T,\mathcal{Y})$ consists in finding a vector $y^\star \in \mathcal{Y}$ such that $T(y^\star)^\top(y - y^\star) \geq 0$ for all $y \in \mathcal{Y}$ and its solution set is denoted by $\text{SOL}(T,\mathcal{Y})$. For a closed set $S \subseteq \mathbb{R}^n$, the mapping $\text{proj}_S : \mathbb{R}^n \rightarrow S$ denotes the projection onto $S$, i.e., $\text{proj}_S(x) \coloneqq \argmin_{y \in S}\|y - x \|$. The set-valued mapping $N_S : \mathbb{R}^n  \rightrightarrows \mathbb{R}^n$ denotes the normal cone operator for the set $S$, i.e., $N_S(x) = \emptyset $ if $x \notin S$ and $N_S(x) = \{v \in \mathbb{R}^{n} \mid \sup_{z \in S} v^\top (z - x) \leq 0 \}$ otherwise. 


\subsubsection{Graph theory}\label{sec:graph}
Let $\mathcal{G} = (\mathcal{N}, \mathcal{E})$ be an undirected graph connecting a set of vertices $\mathcal{N} = \{1,\ldots,N\}$ through a set of edges  $\mathcal{E} \subseteq \mathcal{N} \times \mathcal{N}$, with $|\mathcal{E}| = E$. The unordered pair of vertices $(i, j) \in \mathcal{E}$ if and only if agents
$i$ and $j$ can exchange information. The set of neighbors of
agent $i$ is defined as $\mathcal{N}_i = \{j \in \mc N \mid (i,j) \in \mathcal{E}\}$ and the degree $|\mathcal{N}_i|$ of vertex $i$ corresponds to the cardinality of the set $\mathcal{N}_{i}$. A graph $\mathcal{G}$ is connected if and only if there exists a path between
any two vertices of $\mathcal{G}$. 
We denote by $L \in \mathbb{R}^{N \times N}$ the Laplacian matrix of the
graph $\mathcal{G}$, with $L_{ij} = |\mathcal{N}_i|$ if $i=j$, $L_{ij} = -1$ if $(i,j) \in \mathcal{E}$, $L_{ij} = 0$ otherwise. For an undirected and connected graph, it holds $L = L^\top$.
\section{Generalized Nash equilibrium problems with uncertain coupling constraints}\label{sec:reformulation}
We start by formalizing the multi-agent, uncertain game considered, and then we propose a worst case-based formulation by making use of tools proper of robust optimization.

\subsection{Mathematical setup}\label{sec:math}
We consider an uncertain noncooperative game among $N$ agents, indexed by $i \in \mathcal{I} \coloneqq \{1,\ldots, N\}$, where each agent $i$ makes decisions $x_i \in \mathbb{R}^{n_i}$ in a local constraint set $\Omega_i \coloneqq \{x_i \in \mathbb{R}^{n_i} \mid C_i x_i\leq c_i\}$, for given pairs $\{(C_i, c_i)_{i \in \mc I}\}$ of appropriate dimensions. The vector of collective strategies $\bs{x} \coloneqq \textrm{col}\left((x_i)_{i \in \mathcal{I}}\right) \in \R^n$, $n = \sum_{i\in \mc I} n_i $, is thus constrained to belong to $\Omega \coloneqq \prod_{i \in \mc I} \Omega_i$.
We also denote with $\bs{x}_{-i} \coloneqq \textrm{col}\left((x_j)_{j \in \mc I \setminus \{i\}}\right)$ the vector obtained by stacking all agents' strategies but the $i$-th one. In this context, each agent aims at minimizing a predefined cost function $J_i : \mathbb{R}^n \to \mathbb{R}$ that therefore depends both on its own decision $x_i$, as well as on the decisions of all the other agents $\bs{x}_{-i}$. 

\begin{stassumption}
For each $i \in \mathcal{I}$, $x_i \mapsto J_i(x_i, \bs{x}_{-i})$ is a convex, $\mc C^1$ function, for all $\bs{x}_{-i} \in \R^{n_{-i}}$, while $\Omega_i \neq \emptyset$ is a convex, compact set.
\hfill$\square$
\end{stassumption}

We consider \emph{generalized games} where the agents compete for shared, yet possibly uncertain, resources, thus coupling the agents' decisions also at the feasible set level. Specifically, we consider linear coupling constraints in the form:
\begin{equation}\label{eq:robust_constraint}
    A_i(\delta_i)x_i + \sum_{j \in \mc I \setminus \{i\}}A_j(\delta_j)x_j \leq b(\delta),
\end{equation}
which are affected by the uncertain parameters vector \mbox{$\bs{\delta}\coloneqq \textrm{col}(\textrm{col}((\delta_i)_{i \in \mc I}), \delta) \subseteq \mathbb{R}^{\sum_{i \in \mc{I}}p_i + p}$}, where $\delta_i \in \R^{p_i}$, $\delta \in \R^p$. We hence stress that \eqref{eq:robust_constraint} presents both \textit{local} uncertainties $\{\delta_i\}_{i \in \mathcal{I}}$ affecting the way each agent contributes to the coupling constraints, as well as \textit{global} uncertainties encoded by $\delta$, affecting the vector of shared resources. Formally, we consider a probability space $(\bs{\Delta}, \mathcal{D}, \mathbb{P})$, where $\bs{\Delta} \coloneqq \left( \prod_{i \in \mc I}\Delta_i\right) \times \Delta \subseteq \mathbb{R}^{\sum_{i \in \mc{I}}p_i + p} $ represents the set of values that $\bs{\delta}$ can take, $\mathcal{D}$ is a $\sigma$-algebra and $\mathbb{P}$ is a probability measure over $\mathcal{D}$. Assuming that $\bs{\Delta}$ is bounded, we analyze the coupling constraints in \eqref{eq:robust_constraint} under the lens of robust optimization, thus
requiring it to hold true for all possible (joint) realizations of the uncertain parameter \mbox{$\bs{\delta} \in \bs{\Delta}$.} It follows that \eqref{eq:robust_constraint} amounts to a \textit{robust constraint} and sets $(\{\Delta_i\}_{i \in \mc I}, \Delta)$ are user-prescribed \textit{primitive uncertainty} sets, assumed to be polytopes with the origin in their interiors,
\begin{equation}\label{eq:polyhedral:set}
    \begin{aligned}
    \Delta_i & \coloneqq \{ \delta_i \in \mathbb{R}^{p_i} \mid D_i \delta_i \leq d_i\} \quad \text{for all } \: i \in \mc I,\\
    \Delta & \coloneqq \{ \delta \in \mathbb{R}^{p} \mid D \delta \leq d\},
    \end{aligned}
\end{equation}
for known pairs $\{(D_i, d_i)_{i \in \mc I}\}$ and $(D,d)$, with $D_i \in \mathbb{R}^{m_i \times p_i}, D \in \mathbb{R}^{l \times p}$ and $d_i, d$ sized accordingly.
As common in the robust optimization literature \cite{robust_optimization}, we will make use of the following assumption:
\begin{stassumption}\label{ass:constraintwise}
The uncertain parameter $\bs{\delta} \in \bs{\Delta}$ acts constraint-wise on the inequalities in \eqref{eq:robust_constraint}. 
\hfill$\square$
\end{stassumption}

As a direct consequence, we are hence entitled to study each constraint in \eqref{eq:robust_constraint} separately. Note that Standing Assumption \ref{ass:constraintwise} holds without loss of generality -- see \cite[pp.~11-12]{robust_optimization}.

For a given $\boldsymbol{\delta} \in \boldsymbol{\Delta}$, let  $\mc{X}_{\bs{\delta}} \coloneqq \{\bs{x} \in \mathbb{R}^n \mid A(\bs{\delta}) \bs{x} \leq b(\bs{\delta})\}$ denote the constraint set associated with some realization $\boldsymbol{\delta}$. We thus define the collective feasible set associated to the uncertain GNEP as
$\mathcal{X} \coloneqq \{\cap_{\boldsymbol{\delta} \in \boldsymbol{\Delta}} \mathcal{X}_{\boldsymbol{\delta}} \} \cap \Omega$, which we assume satisfying the following condition:

\begin{stassumption}\label{ass:slater}
The set $\mathcal{X}$ is nonempty and satisfies Slater's constraint qualification.
\hfill$\square$
\end{stassumption}

Standing Assumption~\ref{ass:slater} requires the intersection of all the constraint sets spanned by $\boldsymbol{\delta} \in \boldsymbol{\Delta}$ to be nonempty. Albeit restrictive, note that this condition is equivalently postulated in similar works (see, for example, \cite[Ass.~1]{Filippo}) and it is required for a well-posed problem formulation. 
The uncertain GNEP can hence be described by the following collection of inter-dependent optimization problems:
\begin{equation}\label{eq:game}
\forall i \in \mathcal{I} : \left\{
\begin{aligned}
    & \underset{x_{i} \in \Omega_i}{\textrm{min}} &&  J_{i}\left(x_{i}, \boldsymbol{x}_{-i}\right) \\
    & ~\text { s.t. } && A(\bs{\delta}) \bs{x} \leq b(\bs{\delta}), \: \forall\: \bs{\delta} \in \bs{\Delta}.
\end{aligned}
\right.
\end{equation}

We now introduce a typical tool adopted in game theory and a technical assumption involving it. Specifically, we define the \textit{pseudo-gradient} mapping of the GNEP as:
$$
    F(\bs{x}) \coloneqq \textrm{col}\left( (\nabla_{x_i} J_i(x_i, \bs{x}_{-i}))_{i \in \mc I} \right).
$$
\begin{stassumption}\label{ass:coco}
$F(\bs{x})$ is monotone and $\ell_F$-Lipschitz continuous.
\hfill$\square$
\end{stassumption}

We hence want to compute a Nash equilibrium for \eqref{eq:game} that is valid for all $\bs{\delta} \in \bs{\Delta}$ (namely a worst-case one), according to the following definition:
\begin{definition}\textup{(worst-case generalized Nash equilibrium)}\label{def:rob}
A collective strategy $\bs{x}^\star$ is a worst-case \gls{GNE} of the uncertain game in \eqref{eq:game} if  i) $\bs{x}^\star \in \mc{X}$, and ii) for all $i \in \mc I$,
$$
    J_i(x_i^\star, \bs{x}_{-i}^\star) \leq J_i(x_i, \bs{x}_{-i}^\star),
$$
for any $x_i \in \Omega_i$ so that $(x_i, \bs{x}_{-i}^\star) \in \mathcal{X}$. \hfill$\square$
\end{definition}

Definition \ref{def:rob} enforces $\bs{x}^\star$ to be a \gls{GNE} for all $\bs{\delta} \in \bs{\Delta}$, which is equivalent to enforcing those conditions to hold true for the worst-case disturbance in the uncertainty set. From there the name \textit{worst-case equilibrium}. 
As per standard in algorithmic game theory \cite{Kulkarni}, we however focus on the subclass of \gls{vGNE}. 
For such subclass, the next lemma follows from \cite[Th.~6]{GNEP:Facchinei}:
\begin{lemma}\textup{(Existence of v-GNE)} \label{eq:existence_eq}
The set of v-GNE, as defined in Definition \ref{def:rob}, of the uncertain GNEP in \eqref{eq:game} is nonempty.
\hfill$\square$
\end{lemma}

\subsection{Extended game reformulation}
Finding an equilibrium solution to the uncertain GNEP in \eqref{eq:game} is challenging as it amounts to an infinite-dimensional problem. 
We therefore borrow traditional tools from robust optimization to establish a worst case-based, tractable reformulation of \eqref{eq:game}. In particular, without loss of generality we consider the reformulation of a single \textit{robust} constraint from \eqref{eq:robust_constraint} that we further assume to have the following \textit{affine} dependence from the uncertainty $\bs \delta$:
\begin{equation} \label{eq:single_constraint}
(a_i^0 + P_i \delta_i)^\top x_i +  \sum_{j \in \mc I \setminus \{i\}} (a_j^0 + P_j \delta_j)^\top x_j  \leq b^0 + q^\top \delta.
\end{equation}
This particular instance is considered just to streamline the presentation: the extension to multiple coupling constraints is straightforward under Standing Assumption~\ref{ass:constraintwise}. 

\begin{remark}
While restricting to polyhedral uncertainty sets \eqref{eq:polyhedral:set} and linear coupling constraints \eqref{eq:single_constraint} is limiting, we argue that it is nonetheless a popular choice in the literature with applications, among others, in electric energy systems \cite{robOpt_energy}, transport networks \cite{LIU2018262} and supply chain \cite{Suppy_chain}.
\hfill$\square$
\end{remark}

Next, we reformulate each optimization problem in \eqref{eq:game}  as a deterministic program with finite linear constraints:
\begin{theorem}\label{th:reformulation}
A collective strategy $\bs{x}^\star$ is a worst-case \gls{GNE} of the uncertain GNEP in \eqref{eq:game} if and only if there exist some \mbox{$\bs{y^\star} \in \mathbb{R}^{m}$,} $m = \sum_{i\in \mc I} m_i$, and $z^\star \in \mathbb{R}^l$ such that $(\bs{x}^\star, \bs{y}^\star, z^\star)$ is a GNE of the following \textit{extended} deterministic GNEP:
\begin{equation}\label{eq:game_extended}
\forall i \in \mc I : \left\{
\begin{aligned}
     &\underset{x_{i}, y_{i}, z}{\textrm{min}} &&  J_{i}\left(x_{i}, \boldsymbol{x}_{-i}\right) \\
     &~\textrm{ s.t. } && P_i^\top x_i - D_i^\top y_i = 0,\\
     &&& q + D^\top z = 0,\\
     &&& x_{i}  \in \Omega_i, \, y_{i} \geq 0, \, z \geq 0,\\
     &&& \sum_{i \in \mc I} (a_i^0)^\top x_i \!-\! b^0 \!\leq\! \!- d^\top z \!-\! \sum_{i \in \mc I} d_i^\top y_i.
\end{aligned}
\right.
\end{equation}
\hfill$\square$
\end{theorem}


Some considerations on the nature of the deterministic \textit{extended} game in \eqref{eq:game_extended} are then in order. First, we note that \eqref{eq:game_extended} turns out to be a GNEP, as the last constraint couples the decisions of the agents. Specifically, considering the worst-case on every possible realization of the uncertainties perturbs the feasible set compared to the nominal case, $\sum_{i \in \mc I} (a_i^0)^\top x_i \leq b^0$. Second, while $(x_i, y_i)$ are local variables, $z$ coincides with a global one that, at this stage, precludes the design of a distributed equilibrium seeking algorithm for \eqref{eq:game_extended}. 
In the next section we then propose a fully-distributed GNE seeking algorithm for solving \eqref{eq:game_extended}.

\section{Distributed v-GNE seeking algorithm}\label{sec:algorithm}
We assume that the agents taking part to the uncertain GNEP communicate to each other through a graph $\mathcal{G} = (\mathcal{I}, \mathcal{E})$, where $\mathcal{I}$ is the set of agents and $\mathcal{E}$ is the one of the edges connecting them, $|\mathcal{E}| = E$.  
\begin{stassumption}\label{ass:graph}
    The communication graph $\mathcal{G}$ is undirected and connected.
\end{stassumption}
In addition, we assume a full-decision information setting where the agents know exactly the decisions of the agents influencing their objective function without the need of reconstructing signals.

Thus, to design a fully distributed procedure able to return a solution to the GNEP in \eqref{eq:game_extended}, we propose to endow each agent with a local copy of $z$, i.e., $z_i \in \mathbb{R}^l$, and impose additional constraints enforcing consensus among $z_i$'s, namely $\bar{L}\bs{z} = \bs{0}_{Nl}$, with $\bs{z} = \textrm{col}\left((z_i)_{i \in \mathcal{I}}\right)$ and $\bar{L} = L \otimes I_l$.
For all $i \in \mc I$ we hence define $w_i = \textrm{col}(x_i, y_i, z_i) \in \mathbb{R}^{\eta_i}$, $\eta_i = n_i + m_i + l$. The resulting \textit{extended} game in the variables $w_i$ is a particular instance of the following class of GNEPs:
\begin{equation}\label{eq:game_end}
\forall i \in \mathcal{I} : \left\{
\begin{aligned}
    &\underset{w_i \in \mc W_i}{\textrm{min}} &&  {J}_{i}\left(x_{i}, \boldsymbol{x}_{-i}\right) \\
   &~\text { s.t. } && S_i w_i + \sum_{j\neq i, j \in \mathcal{I}} S_j w_j \leq s, \\
   &&& R_i w_i + \sum_{j\neq i, j \in \mathcal{I}} R_j w_j = \bs{0}, \\
\end{aligned}
\right.
\end{equation}
with $\mc W_i \coloneqq \{w_i \in \R^{\eta_i} \mid H_i w_i \leq h_i, \, G_i w_i = g_i\}$, for matrices $H_i, G_i, S_i, R_i$ and vectors $h_i, g_i, s$ of appropriate dimensions. We denote with $l_{eq} \: (c_{eq})$ and $l_{in} \:(c_{in})$ the total number of local (coupling) equality and local (coupling) inequality constraints in \eqref{eq:game_end}, respectively.

In accordance, the pseudo-gradient characterizing the extended GNEP in \eqref{eq:game_end} turns into
\begin{equation}
    \tilde{F}(\bs{w}) \coloneqq \textrm{col}\left(\left(\nabla_{x_i} J_i(x_i, \bs{x}_{-i}), \bs{0}_{m_i}, \bs{0}_l\right)_{i \in \mc I}\right).
\end{equation}
\begin{proposition}\label{prop:properties}
The extended GNEP in \eqref{eq:game_extended} retains the structural properties of the uncertain GNEP in \eqref{eq:game}, namely
\begin{enumerate}[label=(\roman*)]
\item $\tilde{F}(\bs{w})$ is maximally monotone;
\item The set of \gls{vGNE} of \eqref{eq:game_end} is non-empty.
\hfill$\square$
\end{enumerate}
\end{proposition}



\subsection{An operator splitting approach to v-GNE} 
To solve the GNEP in \eqref{eq:game_end} we start by considering the \gls{KKT} conditions characterizing the optimization problems of the agents, which at optimality read as:
\begin{equation}\label{eq:KKT}
    \begin{cases}
\mathbf{0}_{\eta_i} \in \nabla_{w_{i}} {J}_{i}\left(x_{i}^{\star}, \bs{x}_{-i}^{\star}\right)+S_{i}^{\top} \lambda_i^{\star} + R_i^\top \mu_i^\star +\mathrm{N}_{\tilde{\Omega}_{i}}\left(w_{i}^{\star}\right),\\
\mathbf{0}_{c_{in}} \in \mathrm{N}_{\mathbb{R}_{\geq 0}^{c_{in}}}\left(\lambda_i^{\star}\right)-\left(S \boldsymbol{w}^{*}-s\right),\\
 \mathbf{0}_{c_{eq}} \in -R \bs{w}^\star,
\end{cases}
\end{equation}
where $S = [S_1, \ldots, S_N]$ and $R = [R_1,\ldots, R_N]$. By virtue of the restriction to the subclass of v-GNE, we additionally require $\lambda_1^\star = \lambda_2^\star = \ldots = \lambda_N^\star$ and $\mu_1^\star = \mu_2^\star = \ldots = \mu_N^\star$.

By stacking together the \gls{KKT} conditions in \eqref{eq:KKT}, it is known that the \gls{vGNE} seeking problem for \eqref{eq:game_end} can be
recast as a zero-finding problem involving a suitable set-valued, monotone operator $T$ \cite{Bauschke}. To allow for fully distributed computations, we extend $T$ by endowing each agent with a local copy of the dual variables $\lambda_i$ and $\mu_i$ and driving them towards consensus via the auxiliary variables $v_i$ and $q_i$. The resulting \textit{extended} operator $T$ is thus defined as:
\begin{equation}
    T:\left[\begin{array}{l}
\bs{w} \\
\bs{\nu} \\
\bs{\lambda}\\
\bs{\chi} \\
\bs{\mu}
\end{array}\right] \mapsto\left[\begin{array}{c}
\tilde{F}(\bs{w})+N_{\tilde{\Omega}}(\bs{w})+\bar{S}^{\top} \bs{\lambda} + \hat{R}^\top \bs{\mu} \\
\bar{L} \bs{\lambda}\\
N_{R_{\geq 0}^{Nc_{in} }}(\bs{\lambda})-(\bar{S} \bs{w}-\bar{s}) + \bar{L}(\boldsymbol{\lambda} - \boldsymbol{\nu})\\
\hat{L} \bs{\mu}\\
- \hat{R} \bs{w} + \hat{L} (\bs{\mu} -\bs{\chi})
\end{array}\right],
\end{equation} 
$\bs{\lambda} = \textrm{col}\left((\lambda_i)_{i \in \mathcal{I}}\right)$,  $\bs{\mu} = \textrm{col}\left((\mu_i)_{i \in \mathcal{I}}\right)$,  $\bs{\chi} = \textrm{col}\left((\chi_i)_{i \in \mathcal{I}}\right)$,  $\bs{\nu} = \textrm{col}\left((\nu_i)_{i \in \mathcal{I}}\right)$, $\bar{S} = \text{diag} \left( (S_i)_{i \in \mathcal{I}} \right)$, $\bar{s} = \textrm{col} \left( (s_i)_{i \in \mathcal{I}} \right)$, $\hat{R} = \text{diag} \left( (R_i)_{i \in \mathcal{I}} \right)$, $\bar{L} = L \otimes I_{c_{in}}$, and $\hat{L} = L \otimes I_{c_{eq}}$.

Essentially, the zeros of the mapping $T$ coincide to the variational equilibria of the GNEP \eqref{eq:game_end}, as formalized next.
\begin{proposition}(\hspace{-.03em}\textup{\cite{Bauschke}})\label{prop:zeros_operator}
The collective strategy $\bs{w}^\star$ is a \gls{vGNE} of the game in \eqref{eq:game_end} if and only if there exists $\lambda^\star \in \mathbb{R}^{c_{in}}$ and $\mu^\star \in \mathbb{R}^{c_{eq}}$ such that $\textrm{col}(\bs{w}^\star, \lambda^\star, \mu^\star) \in \text{zer}(T)$. Moreover, if $\textrm{col}(\bs{x}^\star, \lambda^\star, \mu^\star) \in \text{zer}(T)$, then $\bs{w}^\star$ satisfies the KKT conditions in  \eqref{eq:KKT} with $\lambda_i = \lambda^\star, \mu_i \in \mu^\star, \: \forall i \in \mathcal{I}$. 
\hfill$\square$
\end{proposition}

Remarkably, in view of the equivalence between \eqref{eq:game} and \eqref{eq:game_end}, if $\bs{w}^\star \coloneqq \textrm{col}(\bs{x}^\star, \bs{y}^\star, \bs{z}^\star)$ is an equilibrium solution for \eqref{eq:game_end} then $\bs{x}^\star$ is a \gls{vGNE} for the original uncertain game in \eqref{eq:game}. 
Note that the operator $T$ can be split as the sum of two other operators $\mc A$ and $\mc B$. In particular, we have:
$$
\begin{aligned}
\mathcal{A} & \coloneqq \underbrace{\left(\tilde{F}(\bs{w}) \times \bs{0}_{c_{in}} \times \bar{s} \times \bs{0}_{2c_{eq}} \right)}_{\mathcal{A}_1} + \mathcal{A}_2,\\
\mathcal{B} & \coloneqq {\left(N_{\tilde{\Omega}}(\bs{w}) \times \bs{0}_{c_{in}} \times N_{\mathbb{R}^{c_{eq}}_{\geq 0}} \times \bs{0}_{2c_{eq}} \right)},
\end{aligned}
$$
where 
$$
\mathcal{A}_2 : \left[\begin{array}{l}
\bs{w} \\
\bs{\nu} \\
\bs{\lambda}\\
\bs{\chi} \\
\bs{\mu}
\end{array}\right] \mapsto \left[\begin{array}{ccccc}
0 & 0 & \bar{S}^{\top}& 0 & \hat{R}^\top \\
0 & 0 & \bar{L} & 0 & 0 \\
-\bar{S} & -\bar{L} & \bar{L} & 0 & 0 \\
0 & 0 & 0 & 0 & \hat{L}\\
-\hat{R} & 0 & 0 & -\hat{L} & \hat{L}
\end{array}\right]\left[\begin{array}{l}
\bs{w} \\
\bs{\nu} \\
\bs{\lambda}\\
\bs{\chi} \\
\bs{\mu}
\end{array}\right].
$$
\smallskip
\begin{lemma}\label{lem:operator_prop}
The following statements hold true:
\begin{enumerate}[label=(\roman*)]
\item $\mathcal{A}$ is maximally monotone and $\ell_\mathcal{A}$-Lipschitz continuous, with $\ell_\mathcal{A} \coloneqq \ell_F + 4\kappa + |{S}| + |{R}|$, where $\kappa = |L|$;
\item $\mathcal{B}$ is maximally monotone.
\hfill$\square$
\end{enumerate}
\end{lemma}

\subsection{Distributed preconditioned Relaxed-Inertial FBF scheme}
Inspired by \cite{Nesterov,Cui_full}, we design next a fully-distributed relaxed-inertial preconditioned forward-backward-forward (RIpFBF) algorithm to compute a \gls{vGNE} of the GNEP \eqref{eq:game_end} by exploiting the splitting $T = \mathcal{A} + \mathcal{B}$. Specifically, we rely on the following result:
\begin{lemma}{\cite[Prop.~25.26(i)]{Bauschke}}
Given a matrix $\Phi \succ 0$ and a maximally monotone operator $T$, $\Phi^{-1}T$ is maximally monotone w.r.t. the induced norm $\|\cdot\|_{\Phi}$.
\hfill$\square$
\end{lemma}

The main steps of the iterative procedure are reported in Algorithm~\ref{alg:alg1} and follow from the scheme:
\begin{equation}\label{eq:RIPPA}
    \begin{cases}
    Z_k = X_k + \sigma_k (W_k - W_{k-1}),\\
    Y_k =  \textrm{proj}_\Theta \left(\left(\mathrm{Id}-\Phi^{-1} \mathcal{A}\right) (Z_k) \right),\\
    W_{k+1} = (1 - \rho_k) Z_k + \rho_k [Y_k - \Phi^{-1}\left(\mathcal{A}(Y_k)- \mathcal{A}(Z_k)\right)],
    \end{cases}
\end{equation}
where we choose the so-called preconditioning matrix $\Phi$ as
$$
\Phi\coloneqq \text{diag}\left(
\boldsymbol{\alpha}^{-1}, \boldsymbol{\beta}^{-1}, \boldsymbol{\gamma}^{-1}, \boldsymbol{\tau}^{-1}, \boldsymbol{\theta}^{-1}\right),
$$
with $\boldsymbol{\alpha} = \text{diag}((\alpha_i \otimes I_{\mu_i})_{i \in \mathcal{I}})$, $\boldsymbol{\beta} = \text{diag}((\beta_i \otimes I_{c_{in}})_{i=1}^E)$, $\boldsymbol{\gamma} = \text{diag}((\gamma_i \otimes I_{c_{in}})_{i \in \mathcal{I}})$, $\boldsymbol{\tau} = \text{diag}((\tau_i \otimes I_{c_{eq}})_{i=1 }^E)$ and $\boldsymbol{\theta} = \text{diag}((\theta_i \otimes I_{c_{eq}})_{i \in \mathcal{I}})$.
\begin{algorithm}[h!]
\caption{RIpFBF Algorithm}\label{alg:alg1}
\textbf{Initialization}: For all $i \in \mathcal{I}$, set $w_{i,0} \in \mathbb{R}^{\eta_i}$, $\nu_{i,0} \in \boldsymbol{0}_{n_{in}}$ $\lambda_{i,0} \in \mathbb{R}_{\geq 0}^{n_{in}}$ , $\chi_{i,0} \in \boldsymbol{0}_{n_{eq}}$ $\mu_{i,0} \in \mathbb{R}^{n_{eq}}$. 

\textbf{Iteration} ($k \in \mathbb{N}$)\textbf{:} for all $i \in \mathcal{I}$ do

(\texttt{S1}) \textbf{Inertial step:}
$$
\begin{aligned}
w_{i,k}^{in} & =  w_{i,k} + \sigma_k (w_{i,k} - w_{i,k-1}),\\
\nu_{i,k}^{in} & =  \nu_{i,k} + \sigma_k (\nu_{i,k} - \nu_{i,k-1}),\\
\lambda_{i,k}^{in} & =  \lambda_{i,k} + \sigma_k (\lambda_{i,k} - \lambda_{i,k-1}),\\
\chi_{i,k}^{in} & =  \chi_{i,k} + \sigma_k (\chi_{i,k} - \chi_{i,k-1}),\\
\mu_{i,k}^{in} & =  \mu_{i,k} + \sigma_k (\mu_{i,k} - \mu_{i,k-1}).\\
\end{aligned}
$$

(\texttt{S2}) \textbf{Forward-Backward step:} from $j \in \mathcal{N}_i$ get $\mu_{j,k}^{in} = [w_{j,k}^{in}, v_{j,k}^{in}, \lambda_{j,k}^{in}, q_{j,k}^{in}, \mu_{j,k}^{in}]$ and update
$$
\begin{aligned}
w_{i,k}^{pr}  & = \text{proj}_{\tilde{\Omega}_i} \left( w_{i,k}^{in} \!-\! \alpha_i \left( \tilde{F}_i(\bs{w}_k^{in}) \!+\! S_i^\top \lambda_{i,k}^{in} \!+\! R_i^\top \mu_{i,k}^{in}\right) \right) \\
\nu_{i,k}^{pr} & = \nu_{i,k}^{in} + \beta_i(|\mathcal{N}_i| \lambda_{i,k}^{in} - \textstyle\sum_{j \in \mathcal{N}_i} \lambda_{j,k}^{in}),\\
\lambda_{i,k}^{pr} & = \text{proj}_{\mathbb{R}_{\geq 0}^{c_{in}}}  \left(\lambda_{i,k}^{in} + \gamma_i\left(S_i w_{i,k}^{in}  - s_i -  \nu_{i,k}^{in}) \right) \right),\\
\chi_{i,k}^{pr} & = \chi_{i,k}^{in} + \tau_i(|\mathcal{N}_i| \mu_{i,k}^{in} - \textstyle\sum_{j \in \mathcal{N}_i} \mu_{j,k}^{in}),\\
\mu_{i,k}^{pr} & =  \mu_{i,k}^{in} + \theta_i\left(R_i w_{i,k}^{in} -\chi_{i,k}^{in}) \right).
\end{aligned}
$$
(\texttt{S3}) \textbf{Relaxed step:} 
$$
\begin{aligned}
w_{i,k+1} & =  (1 \!-\! \rho_k) w_{i,k}^{in} \!+\! \rho_k \Big[w_{i,k}^{pr} \!-\! \alpha_i\left( \tilde{F}_i(w_{k}^{pr}) \!-\! \tilde{F}_i(w_k^{in}) \right.\\
& \left. + S_i^\top (\lambda_{i,k}^{pr} - \lambda_{i,k}^{in}) + R_i^\top  (\mu_{i,k}^{pr}-\mu_{i,k}^{in}) \right)\Big],\\
\nu_{i,k+1} & =  (1 - \rho_k) \nu_{i,k}^{in} + \rho_k \Big[\nu_{i,k}^{pr} - \beta_i \Big( (|\mathcal{N}_i|\lambda_{i,k}^{pr} \\ 
& - \textstyle\sum_{j \in \mathcal{N}_i} \lambda_{j,k}^{pr}) 
- (|\mathcal{N}_i|\lambda_{i,k}^{in} - \textstyle\sum_{j \in \mathcal{N}_i} \lambda_{j,k}^{in} ) \Big)\Big],\\
\lambda_{i,k+1} & =  (1 - \rho_k) \lambda_{i,k}^{in} + \rho_k \Big[ \lambda_{i,k}^{pr} -\gamma_i\Big(-S_i(w_{i,k}^{pr} - w_{i,k}^{in}) \\
&  + |\mathcal{N}_i|(\lambda_{i,k}^{pr} - \lambda_{i,k}^{in}) - |\mathcal{N}_i|(\nu_{i,k}^{pr} - \nu_{i,k}^{in}) \\
& - (\textstyle\sum_{j \in \mathcal{N}_i} \lambda_{j,k}^{pr} - \textstyle\sum_{j \in \mathcal{N}_i} \lambda_{j,k}^{in}) \\
& +  (\textstyle\sum_{j \in \mathcal{N}_i} \nu_{j,k}^{pr} - \textstyle\sum_{j \in \mathcal{N}_i} \nu_{j,k}^{in}) \Big) \Big],\\
\chi_{i,k+1} & =   (1 - \rho_k) \chi_{i,k}^{in} + \rho_k \Big[\chi_{i,k}^{pr} - \tau_i \Big( (|\mathcal{N}_i|\mu_{i,k}^{pr} \\ 
& - \textstyle\sum_{j \in \mathcal{N}_i} \mu_{j,k}^{pr}) 
- (|\mathcal{N}_i|\mu_{i,k}^{in} - \textstyle\sum_{j \in \mathcal{N}_i} \mu_{j,k}^{in} ) \Big)\Big],\\
\mu_{i,k+1} & =  (1 - \rho_k) \mu_{i,k}^{in} + \rho_k \Big[ \mu_{i,k}^{pr} - \theta_i\left(-R_i(w_{i,k}^{pr} - w_{i,k}^{in}) \right.\\
&  + |\mathcal{N}_i|(\mu_{i,k}^{pr} - \mu_{i,k}^{in}) - |\mathcal{N}_i|(\chi_{i,k}^{pr} - \chi_{i,k}^{in}) \\
& - (\textstyle\sum_{j \in \mathcal{N}_i} \mu_{j,k}^{pr} - \textstyle\sum_{j \in \mathcal{N}_i} \mu_{j,k}^{in}) \\
& +  (\textstyle\sum_{j \in \mathcal{N}_i} \chi_{j,k}^{pr} - \textstyle\sum_{j \in \mathcal{N}_i} \chi_{j,k}^{in}) \Big) \Big].
\end{aligned}
$$
\end{algorithm}
We now describe the steps in \eqref{eq:RIPPA}:
\begin{enumerate}
    \item \textit{Inertial} step: this step penalizes changes from the previously computed decisions in the primal-dual space.
    \item \textit{Forward-Backward} step: each agent updates its strategy through a gradient-based step followed by a consensus-enforcing update of the dual variables estimates, and a dual update in the spirit of Lagrangian methods.
    \item \textit{Relaxed} step: the last step consists on a weighted average between the inertial update and a forward-backward-forward update.
\end{enumerate}

Note that for $\sigma_k = 0$ and $\rho_k = 1$, Algorithm 1 reduces to the popular Tseng's extragradient method \cite{Bauschke}.
The next result finally characterizes the convergence of Algorithm 1 to a \gls{vGNE} of the uncertain GNEP \eqref{eq:game}:
\begin{theorem}\label{th:convergence}
Let $\lambda_{\textrm{min}}(\Phi) \in \left( 0, 1/\ell_\mathcal{A}\right)$, and choose $0 < \sigma_k \leq \bar{\sigma} < 1$ and $\rho_k = 2( 1 - \bar{\sigma})^2/{(1+\ell_\Phi) (2\sigma_k^2 - \sigma_k + 1)}$, with $\ell_\Phi \coloneqq \ell_{\mathcal{A}}/{\lambda_{\textrm{min}}(\Phi)}$. Then, the sequence $\{\bs{w}_k\}_{k \in \N}$ generated by Algorithm 1 converges to some $\bs{w}^\star$, whose subvector $\bs{x}^\star$ is a \gls{vGNE} of the uncertain game \eqref{eq:game}.
\hfill$\square$
\end{theorem}
 \smallskip

\section{Illustrative example}\label{sec:numerics}
We test the effectiveness of our formulations on a GNEP with $N = 5$ agents. For $i \in \mathcal{I} = \{1,\ldots,5\}$, let $x_i^j$ denote the $j$-th component of the strategy of the $i$-th agent. We consider $j = \{1,2\}$, i.e., $x_i \in \mathbb{R}^2$ so that $x_{i} \in [-5,15]^2$. The cost function of the $i$-th agent is
$$
J_i(x_i, \boldsymbol{x}_{-i}) = \tfrac{1}{2}x_i^\top x_i + \tfrac{1}{|\mathcal{N}_i|}\sum_{j \in \mathcal{N}_i}x_i^\top x_j - \alpha_i^\top x_i,
$$
where $\alpha_i = 10(i-1)\boldsymbol{1}_2$. Agents need to meet a coupling constraint as in \eqref{eq:single_constraint},
where $a_i^0 = \begin{bmatrix} 1 & 1\end{bmatrix}, P_i = \begin{bmatrix} 1 & 1\end{bmatrix}^\top, \delta_i \in [-1,1] \subseteq \mathbb{R}, b^0 = 75, Q = 1, \delta \in [-10,10] \subseteq \mathbb{R}$.

\begin{figure}[t!]
    \centering
    \includegraphics[trim={0.5cm 4cm 0.5cm 4cm},clip,width=\linewidth]{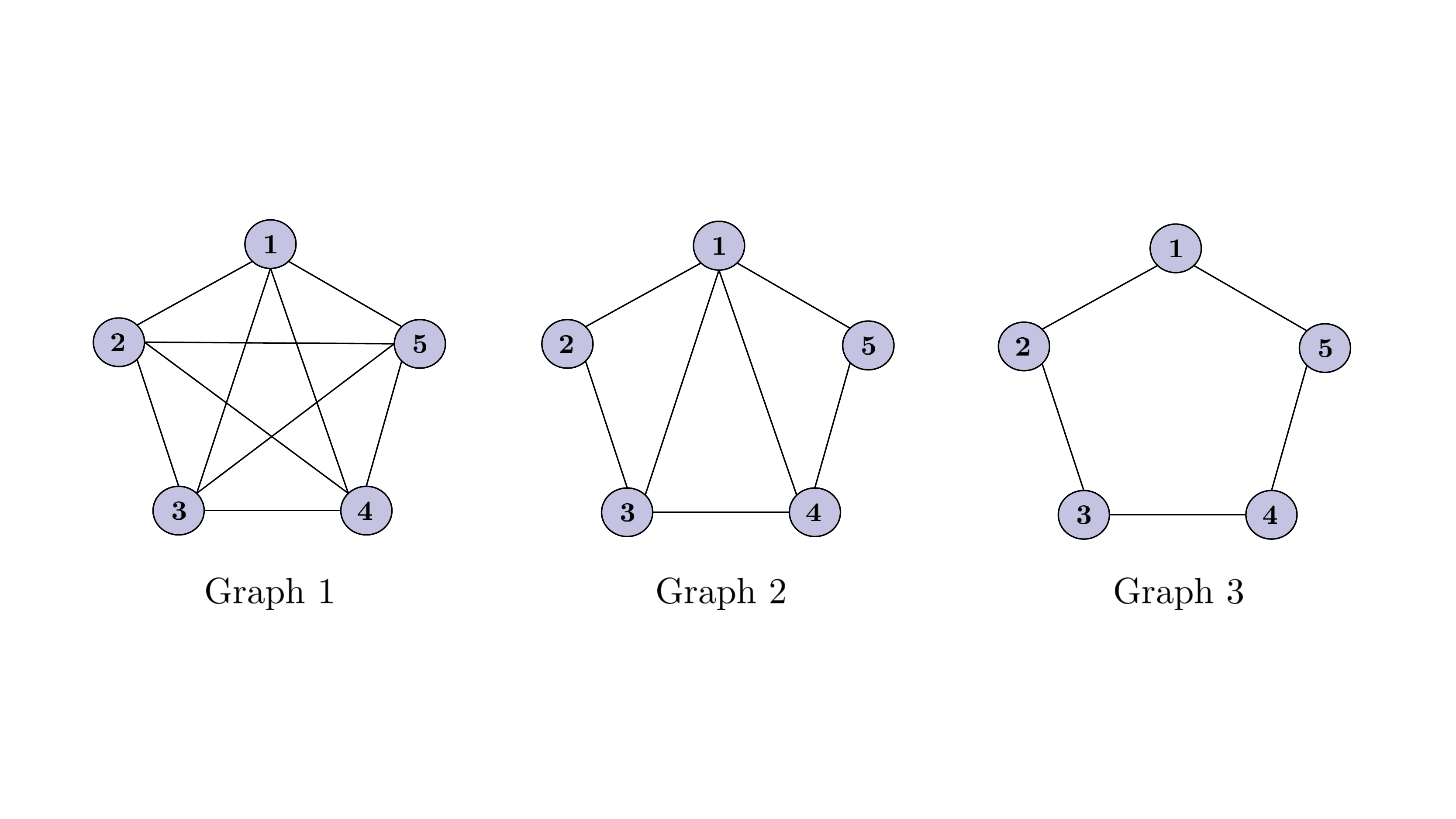}
    \caption{Graph topologies considered in the illustrative example.}
    \label{fig:graphs}
\end{figure}

\begin{figure}[t!]
    \centering
    \includegraphics[width=\linewidth]
    {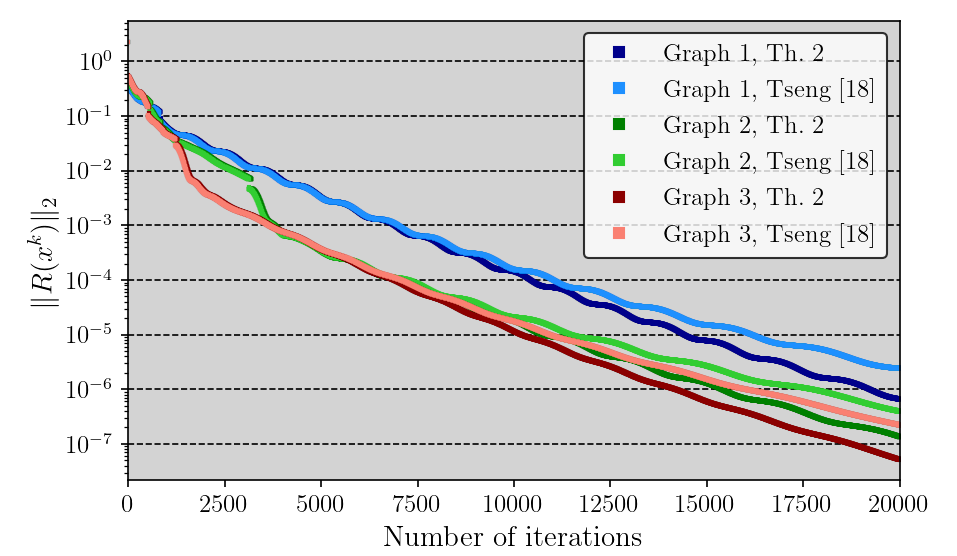}
    \caption{Convergence plot for Algorithm 1 for different problem setups.}
    \label{fig:convergence_base}
\end{figure}

\begin{figure}[t!]
    \centering
    \includegraphics[width=\linewidth]
    {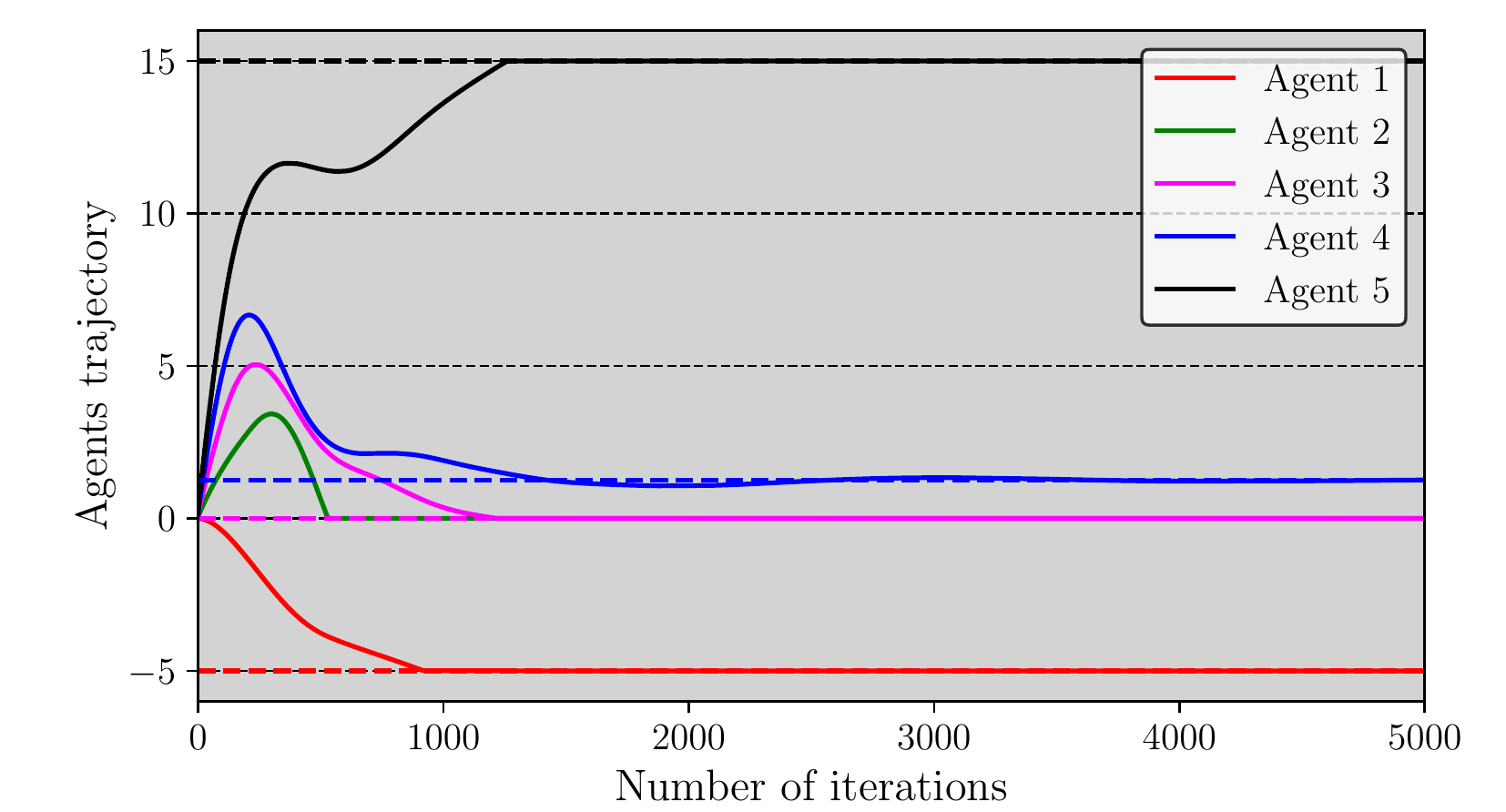}
    \caption{Agents trajectory for $x_1$ ($x_2$ is the same) computed with Algorithm 1 (solid lines) and a centralized solver (dotted lines) for simulation with Graph 3. The $x$-axis is cut at 5000 iterations for readability.}
    \label{fig:traj}
\end{figure}

We consider three different graph topologies with decreasing connectivity, as reported in Fig.~\ref{fig:graphs}. 
We employ Algorithm 1 with inertial parameter $\sigma_k = \bar{\sigma}(1 - \frac{1}{k+1})$ and relaxation parameter $\rho_k$ chosen according to Theorem \ref{th:convergence}. The preconditioning matrix is populated with step sizes that are progressively decreased in the interval $(0, \frac{1}{\ell_\mathcal{A}})$ with even spaces to achieve a better trade-off between convergence rate and accuracy. Figure \ref{fig:convergence_base} shows the convergence of the scheme towards an equilibrium of the original uncertain GNEP \eqref{eq:game} for different tuning parameter configurations and graph topologies. Convergence is monitored according to the natural residual \cite[p.~22]{residual}:
$$
R(x_k) = x_k - (\mathcal{B} \circ (Id - \phi^{-1}\mathcal{A})(x_k)).
$$
We note that the convergence rate is strongly influenced by the choice of the hyperparameters and the graph topology. In general, we observe that the accelerated scheme from Algorithm~1 shows better convergence properties than the (non-accelerated) Tseng extragradient method regardless of the graph topology. Additionally, we find that the ring graph shows the best convergence rate: while this seems counterintuitive at first, as more densely connected graph should allow for an easier information exchange among agents, we remark that the graph topology also affects the Lipschitz constant of $\mathcal{A}$, generally requiring smaller step sizes for highly connected graphs according to the (conservative) bounds in Theorem 2. For completeness, we report Fig~\ref{fig:traj} that compares the trajectories of the agents computed with the (fully-distributed) Algorithm~1 and a centralized solver.

\section{Conclusion}\label{sec:end}
We have presented a worst case-based reformulation of a GNEP with uncertain coupling constraints, along with an algorithm to solve the resulting deterministic, extended game. By exploiting the specific structure of those constraints and the way the uncertainty affects them, we have adopted tools from robust optimization to derive an equivalent, yet tractable, extended GNEP, which is shown to retain all the properties of the original game. By relying on monotone operator theory, we have successively derived an iterative, fully-distributed accelerated algorithm to compute the Nash equilibria of the extended game in the worst-case. 


\appendix
\subsection{Proofs of \S \ref{sec:reformulation}}\label{subsec:app_A}

\textit{Proof of Lemma \ref{eq:existence_eq}}:
The proof follows from \cite[Th.~6]{GNEP:Facchinei} after observing that \text{(i)} $\mc X$ is a convex, compact and non-empty set that satisfies Slater's constraint qualification in view of Standing Assumption 3; and (ii) $F(\bs{x})$ is monotone and $\ell_F$-Lipschitz continuous according to Standing Assumption~\ref{ass:coco}, thus also maximally monotone \cite[Cor.~20.25]{Bauschke}.
$\hfill \blacksquare$

\smallskip

\textit{Proof of Theorem \ref{th:reformulation}}:
To retrieve the worst-case relation in \eqref{eq:single_constraint}, we consider the $i$-th agent for which we have to compute
\begin{equation}\label{eq:primal}
\textrm{max}_{\delta_i \in \Delta_i} (a_i^0 + P_i \delta_i)^\top x_i = (a_i^0)^\top x_i + \textrm{max}_{\delta_i \in \Delta_i} \delta_i^\top P_i^\top x_i ,
\end{equation}
which corresponds to a (manifestly feasible) \gls{LP}, and therefore the associated dual problem amounts to:
\begin{equation}\label{eq:RO_min}
\left\{
\begin{aligned}
    & \underset{y_i \in \mathbb{R}^{m_i}_{\geq 0}}{\textrm{min}} && d_i^\top y_i\\
    & ~\text{ s.t.} && P_i^\top x_i - D_i^\top y_i = 0, 
\end{aligned}
\right.
\end{equation}
with vector of Lagrange multipliers $y_i$. Similarly, we can reformulate the worst-case condition characterizing the RHS in \eqref{eq:single_constraint}, i.e., $ \textrm{min}_{\delta \in \Delta} b^0 + q^\top \delta = b^0 + \textrm{min}_{\delta \in \Delta} q^\top \delta$, to obtain
\begin{equation} \label{eq:RO_max}
\left\{
\begin{aligned}
    & \underset{z \in \mathbb{R}^{l}_{\geq 0}}{\textrm{max}} && -d^\top z\\
    & ~\text{ s.t.} && \quad q + D^\top z = 0.
\end{aligned}
\right.
\end{equation}
At this point, we can omit the minimization term from \eqref{eq:RO_min} (resp. the maximization term from \eqref{eq:RO_max}) since it is sufficient that the constraint holds for at least one $z_i$ (resp. $z$). The result hence follows by adopting algebraic manipulations to rearrange terms. $\hfill \blacksquare$

\subsection{Proofs of \S \ref{sec:algorithm}}\label{subsec:app_B}
\textit{Proof of Proposition \ref{prop:properties}}:
(i) We start by showing that, also in this case, $\tilde{F}(\bs{w})$ is monotone and $\ell_F$-Lipschitz continuous. We have: 
$$
\begin{aligned}
\langle \tilde{F}(\bs{w}_1) - \tilde{F}(\bs{w}_2), \bs{w}_1 - \bs{w}_2 \rangle &= \langle F(\bs{x}_1) - F(\bs{x}_2)\rangle \geq 0,
\end{aligned}
$$
in view of the monotonicity of $F(\bs{x})$. Moreover,
$$
\begin{aligned}
    \|\tilde{F}(\bs{w}_1) &- \tilde{F}(\bs{w}_2)\| = \|F(\bs{x}_1) - F(\bs{x}_2)\| \\
    & \leq  \ell_F \|\bs{x}_1 - \bs{x}_2\| \\
    & \leq \ell_F \left(\|\bs{x}_1 - \bs{x}_2\| +  \|\bs{y}_1 - \bs{y}_2\| + \|z_1 - z_2\|  \right),
\end{aligned}
$$
as $F(\bs{x})$ is $\ell_F$-Lipschitz continuous. Observing that the feasible set of the GNEP \eqref{eq:game_extended} is compact concludes the proof.\\
(ii) This part follows similarly to the proof of Lemma 1. 
$\hfill \blacksquare$

\smallskip

\textit{Proof of Lemma \ref{lem:operator_prop}}:
(i) Note that $\mathcal{A}_1$ is maximally monotone since it is obtained by stacking $\tilde{F}(\bs{w})$, that it maximally monotone in view of Proposition~\ref{prop:properties}, and a constant. Additionally, it is easy to see that it is $\ell_F$-Lipschitz continuous. Next, note that $\mathcal{A}_2$ is a skew-symmetric matrix since $L = L^\top$ as a consequence of Standing Assumption 5, thus it is maximally monotone \cite[Ex.~20.30]{Bauschke}. Then, $\mathcal{A}$ is maximally monotone as sum of two maximally monotone operators \cite[Prop.~21.24]{Bauschke}. Moreover, following a similar line of proof as in \cite[Lemma~5]{Pavel}, $\mathcal{A}_2$ is $(|S| + |R| + 4\kappa)$-Lipschitz continuous. Hence, $\mathcal{A}$ is $\ell_\mathcal{A} \coloneqq \ell_F + 4\kappa + |\bar{S}| + |\hat{R}|$-Lipschitz continuous \cite{Bauschke}.\\
(ii) The mapping $\mathcal{B}$ is constructed by means of normal cones of closed non-empty convex sets, thus it is maximally monotone \cite[Prop~21.23]{Bauschke}.
$\hfill \blacksquare$

\smallskip

\textit{Proof of Theorem \ref{th:convergence}}:
We take inspiration from \cite{Cui_full}, \cite{Cui} to show that the claim holds true. In particular, we first derive the fundamental recursion between successive iterates of Algorithm~\ref{alg:alg1}, and then show the recursion enjoys a Lyapunov-like decrease ensuring convergence.
Let $T_k \coloneqq Y_k - \Phi^{-1}(\mathcal{A}(Y_k) - \mathcal{A}(Z_k))$ such that $W_{k+1} = (1-\rho_k)Z_k + \rho_k T_k$. Recall that $\text{zer}(\mathcal{A} + \mathcal{B}) \neq \emptyset$ in view of Proposition \ref{prop:properties} and \ref{prop:zeros_operator}. For any $\omega^\star \in \text{zer}(\mathcal{A} + \mathcal{B})$, we have that:
\small{$$
\begin{aligned}
\left\|Z_{k}-\omega^\star\right\|_\Phi^{2} &=\left\|Z_{k}-Y_{k}+Y_{k}-T_{k}+T_{k}-\omega^\star\right\|_\Phi^{2} \\
&=\left\|Z_{k}-Y_{k}\right\|_\Phi^{2}-\left\|Y_{k}-T_{k}\right\|_\Phi^{2}+\left\|T_{k}-\omega^\star\right\|_\Phi^{2} \\
&{+2\left\langle Z_{k}-Y_{k}, Y_{k}-\omega^\star\right\rangle_\Phi +2\left\langle Y_{k}-T_{k}, T_{k}-\omega^\star\right\rangle_\Phi}\\
&=\left\|Z_{k}-Y_{k}\right\|_\Phi^{2}-\left\|Y_{k}-T_{k}\right\|_\Phi^{2}+\left\|T_{k}-\omega^\star\right\|_\Phi^{2} \\
&+2\left\langle Z_{k}-T_{k}, Y_{k}-\omega^\star\right\rangle_\Phi\\
\end{aligned}
$$}
where
$$
\begin{aligned}
\left\|Y_{k}-T_{k}\right\|_\Phi^{2} &=\left\|\Phi^{-1}\left(\mathcal{A}(Y_k) - \mathcal{A}(Z_k)\right)\right\|_\Phi^{2} \\
&\leq |\Phi^{-1}| \left\|\left(\mathcal{A}(Y_k) - \mathcal{A}(Z_k)\right)\right\|^{2}\\
&\leq |\Phi^{-1}|\ell_{\mathcal{A}}^2 \left\|Y_k-Z_k\right\|^{2}\\
& \leq \tfrac{\lambda_{\textrm{max}}(\Phi^{-1})}{\lambda_{\textrm{min}}(\Phi)}\ell_{\mathcal{A}}^2\left\|Y_{k}-Z_{k}\right\|_\Phi^{2}\\
& = \left(\tfrac{\ell_{\mathcal{A}}}{\lambda_{\textrm{min}}(\Phi)}\right)^2\left\|Y_{k}-Z_{k}\right\|^{2}
\end{aligned}
$$
and
$\left\langle Z_{k}-T_{k}, Y_{k}-\omega^\star\right\rangle\geq 0$ from the monotonicity of $\mathcal{A}$. Then, by defining $\ell_\Phi \coloneqq \ell_{\mathcal{A}}/\lambda_{\textrm{min}}(\Phi)$, we have:
\small{\begin{equation}
\begin{aligned}
\left\|Z_{k}-\omega^\star\right\|_\Phi^{2} {\geq}\left\|Z_{k}-Y_{k}\right\|_\Phi^{2}- \ell_{\Phi}^2\left\|Y_{k}-Z_{k}\right\|_\Phi^{2}+\left\|T_{k}-\omega^\star\right\|_\Phi^{2},
\end{aligned}
\end{equation}}\normalsize
which leads to
\begin{equation}\label{eq:Tomega}
    \|T_k - \omega^\star\|_\Phi^2 \leq \|Z_k - \omega^\star)\|_\Phi^2 - (1 - \ell_{\Phi}^2) \|Y_k - Z_k\|_\Phi^2.
\end{equation}
From \eqref{eq:Tomega} we immediately obtain:
\begin{equation}
\begin{aligned}
&\left\|W_{k+1}-\omega^\star\right\|_\Phi^{2}\\
=& \left\|\left(1-\rho_{k}\right) Z_{k}+\rho_{k} T_{k}-\omega^\star\right\|_\Phi^{2} \\
=&\left(1-\rho_{k}\right)\left\|Z_{k}-\omega^\star\right\|_\Phi^{2}+\rho_{k}\left\|T_{k}-\omega^\star\right\|_\Phi^{2} \\
& \quad -\rho_{k}\left(1-\rho_{k}\right)\left\|T_{k}-Z_{k}\right\|_\Phi^{2} \\
\leq  & \left(1-\rho_{k}\right)\left\|Z_{k}-\omega^\star\right\|_\Phi^{2}+\rho_{k}\left\|Z_{k}-\omega^\star\right\|_\Phi^{2}\\
& \quad - \rho_k(1 -\ell_{\Phi}^2)\|Y_k - Z_k\|_\Phi^2 -\tfrac{1-\rho_{k}}{\rho_{k}}\left\|W_{k+1}-Z_{k}\right\|_\Phi^{2}\\
= & \|Z_k - \omega^\star\|_\Phi^2 - \rho_k(1 - \ell_{\Phi}^2)\|Y_k - Z_k\|_\Phi^2\\
&~\quad-\tfrac{1-\rho_{k}}{\rho_{k}}\left\|W_{k+1}-Z_{k}\right\|_\Phi^{2}.
\end{aligned}
\end{equation}
Additionally, note that
\begin{equation}
\begin{aligned}
\tfrac{1}{\rho_{k}}&\left\|W_{k+1}-Z_{k}\right\|_{\Phi} =\left\|T_{k}-Z_{k}\right\|_{\Phi} \\
& = \left\|(\mathcal{A}(Y_k) - \mathcal{A}(Z_k))\right\|_{\Phi^{-1}}+\left\|Y_{k}-Z_{k}\right\|_{\Phi} \\
& \leq\left(1+\ell_\Phi\right)\left\|Y_{k}-Z_{k}\right\|_{\Phi},
\end{aligned}
\end{equation}
that is 
$$-\tfrac{1}{\rho_k^2 (1 + \ell_\Phi)^2} \|W_{k+1} - Z_k\|_{\Phi}^2 \geq - \|Y_k - Z_k\|_{\Phi}^2,$$
and by multiplying both sides for $\rho_k(1 - \ell_{\Phi}^2)$ we obtain 
$$-\tfrac{(1 - \ell_{\Phi}^2)}{\rho_k (1 + \ell_\Phi)^2} \|W_{k+1} - Z_k\|_{\Phi}^2 \geq - \rho_k(1 - \ell_{\Phi}^2)\|Y_k - Z_k\|_{\Phi}^2.$$
Plugging-in the latter expression in (17), we get
\begin{equation}
\begin{aligned}
& \|W_{k+1} - \omega^\star\|_{\Phi}^2 \\
& \leq \|Z_k - \omega^\star\|_{\Phi}^2  - \left( \tfrac{1 - \rho_k}{\rho_k} + \tfrac{(1 - \ell_\Phi^2)}{\rho_k(1 + \ell_\Phi)^2}  \right)\|W_{k+1} - Z_k\|_{\Phi}^2\\
& = \|Z_k - \omega^\star\|_{\Phi}^2 - \left( \tfrac{2}{\rho_k(1 + \ell_\Phi)} -1 \right)\|W_{k+1} - Z_k\|_{\Phi}^2.
\end{aligned}
\end{equation}
Next, we turn our attention to the RHS of (19). By leveraging the definitions in (15) we see that:
\begin{equation}
\begin{aligned}
\|W_{k+1} - Z_{k}\|_{\Phi}^2 & \geq (1 - \sigma_k)\|W_{k+1} - W_k\|_{\Phi}^2 \\
& + (\sigma_k^2 - \sigma_k)\|_{\Phi}W_k - W_{k-1}\|_{\Phi}^2
\end{aligned}
\end{equation}

\begin{equation}
\begin{aligned}
\|Z_{k} - \omega^\star\|_{\Phi}^2 & = (1 + \sigma_k) \|W_k - \omega^\star\|_{\Phi}^2 - \sigma_k\|W_{k-1} - \omega^\star\|_{\Phi}^2 \\
& + \sigma_k(1 + \sigma_k)\|W_k - W_{k-1}\|_\Phi^2.
\end{aligned}
\end{equation}
Combining (19)--(21) and rearranging the terms yields:
\begin{equation}
\begin{aligned}
& \|W_{k+1} - \omega^\star\|_{\Phi}^2 -\sigma_k \|W_k - \omega^\star\|_{\Phi}^2 + \\
& (1-\sigma_k)\left( \tfrac{2}{\rho_k(1 + \ell_\Phi)} -1 \right) \|W_{k+1} - W_k\|_{\Phi}^2 \\
& \leq \|W_k - \omega^\star\|_{\Phi}^2 - \sigma_k \|W_{k-1} - \omega^\star\|_{\Phi}^2 + \\
& \Bigg[\sigma_k(1 + \sigma_k) - (\sigma_k^2 - \sigma_k)\left( \tfrac{2}{\rho_k(1 + \ell_\Phi)} -1 \right) \Bigg] \|W_{k} - W_{k-1}\|_{\Phi}^2.
\end{aligned}
\end{equation}

We now choose the sequences $\{\sigma_k\}_{k \in \N}$ and $\{\rho_k\}_{k \in \N}$ as $ 0 < \sigma_k \leq \bar{\sigma} < 1$ and $\rho_k = \tfrac{2(1 - \bar{\sigma}^2)}{(1+ \ell_\Phi) (2 \sigma_k^2 - \sigma_k + 1)}$ and show that this leads to the sought Lyapunov-like decrease across the iterates. To show this, note that under this choice the coefficient multiplying the term $\|W_k - W_{k-1}\|^2$ is
$$
\begin{aligned}
&\sigma_k(1 + \sigma_k) + (\sigma_k^2 - \sigma_k) \left( \tfrac{2\sigma_k^2 - \sigma_k +1}{(1 - \sigma_k)^2} \right)\\
&= -\tfrac{\sigma_k(2\sigma_k^2 - \sigma_k +1)}{1 - \sigma_k} \leq 0.
\end{aligned}
$$
On the contrary, the right-hand side in (23) is non-positive under the given parameters tuning. Therefore, if we define $$
\begin{aligned}
&H_k(\omega^\star) = \|W_{k} - \omega^\star\|_\Phi^2 - \sigma_k\|W_{k-1} - \omega^\star\|_\Phi^2 \\
&+ (1-\sigma_k)^2\left(\tfrac{2\sigma_k^2 - \sigma_k +1}{(1-\sigma_k)^2}  \right)\|W_{k} - W_{k-1}\|_\Phi^2
\end{aligned}
$$
and $\Delta = - \left[\sigma_k(1 + \sigma_k) + (\sigma_k^2 - \sigma_k) \left( \tfrac{2\sigma_k^2 - \sigma_k +1}{(1 - \sigma_k)^2} \right)\right]$, we obtain the following Lyapunov-like decrease condition
\begin{equation}
    H_{k+1}(\omega^\star) - H_k(\omega^\star) \leq -\Delta \|W_{k+1} - W_{k}\|_\Phi^2,
\end{equation}
which ensure convergence to a fixed point of (14). According to Lemma 3, this corresponds to a zero of $T$, which in turn amounts to a \gls{vGNE} of the extended GNEP in \eqref{eq:game_extended} by Proposition 2. Invoking Theorem 1 concludes the proof.
$\hfill \blacksquare$


\bibliographystyle{IEEEtran}
 \bibliography{biblio.bib}
 
 \end{document}